\documentstyle[preprint,prl,aps]{revtex}
\begin{document}
\draft
\title{PERTURBATION AND VARIATIONAL METHODS IN
NONEXTENSIVE TSALLIS STATISTICS}
\author{E. K. Lenzi$^{1}$, L. C. Malacarne$^{2,3}$, 
and  R. S. Mendes$^{1,3}$ }
\address{$^1$Centro Brasileiro de Pesquisas F\'\i sicas, 
R. Dr.  Xavier Sigaud 150, \\22290-180 Rio de Janeiro-RJ, 
Brazil\\ $^2$Departamento de F\'{\i}sica Matem\'atica, 
Instituto de F\'{\i}sica da Universidade de S\~ao Paulo, \\
Caixa Postal 20516, 01498, S\~ao Paulo-SP, Brazil \\
$^3$Departamento de F\'\i sica, Universidade Estadual de 
Maring\'a, \\
Av. Colombo 5790, 87020-900 Maring\'a-PR, Brazil}
\date{\today }
\maketitle
\begin{abstract}

A unified presentation of the perturbation and variational 
methods for the generalized statistical mechanics based on 
Tsallis entropy is given here. 
In the case of the  variational method, the Bogoliubov 
inequality is generalized in a very natural way following the 
Feynman proof for  the usual statistical mechanics. The inequality 
turns out to be {\it form-invariant } with respect to the entropic 
index $q$. 
The method is illustrated with a simple example 
in classical mechanics. The formalisms developed here 
are expected to be useful in the discussion of nonextensive systems. 
\end{abstract}
\date{\today}


Nonextensive effects are common in many 
branches of the physics, for instance, anomalous 
diffusion\cite{1a,1b,1c,1d}, 
astrophysics with long-range (gravitational) 
interactions\cite{2,2a,2b,2c,2d}, 
some magnetic systems\cite{3,3a,3b}, some surface 
tension questions\cite{4,4a}. 
These examples indicate that the
standard statistical mechanics and thermodynamics 
need some extensions.
In this direction, a theoretical tool based in a 
nonextensive entropy (Tsallis entropy)\cite{T1} has 
successfully been applied in several 
situations, for example, L\'{e}vy-type anomalous 
superdiffusion\cite{6}, 
Euler turbulence\cite{7}, self-gravitating
 systems\cite{7,8,8a,8b,8c}, cosmic background
radiation\cite{9}, peculiar velocities in galaxies\cite{10a},
linear response theory\cite{10b} and eletron-phonon
 interaction\cite{10c},   
and ferrofluid-like systems\cite{11}. 
Lavenda and coworkers\cite{L}  stress that 
any newly proposed entropy\cite{L2} must have 
``concavity" property for it to be correct.
Tsallis\cite{T1} and Mendes\cite{M} have shown that the Tsallis 
entropy indeed satisfies this
criterion and hence meets this requirement of concavity.
The above features of Tsallis entropy thus make it unique 
among other
forms for entropy suggested in the literature.
In this context, it is very important to understand more
 deeply the 
properties of the generalized 
statistical mechanics based on the Tsallis entropy. 
In particular, a generalization of the approximate methods 
of calculation of its thermodynamical functions is of  great
 value, as for instance, 
the semiclassical 
approximation\cite{EMM}, perturbation  and variational methods. 
The present work deals  with the last two questions. 
We develop here the perturbation and variational 
methods in a unified way. 
This approach provides the generalization of the Feynman 
proof\cite{F1} 
of the Bogoliubov inequality,  
which appears to be a natural 
generalization of  this inequality. 
Indeed, we shall prove that
{\it the original form of the inequality is preserved } 
(see inequality (\ref{F13})). 
This inequality does not coincide from that proposed 
in ref.\cite{PT}, 
except for $q=1$. 
This is due to the fact that we use different mathematical 
inequalities  to derive our final results.

The Tsallis entropy and a $q$-expectation 
value for an observable $A$\cite{CT}
is defined respectively as 
$S_q= k \, Tr \,  \rho \left( 1-\rho ^{q-1}\right) /(q-1)$
and $\langle A\rangle _q=Tr\rho^q A$, where $\rho $ 
is the density matrix,  
$q\in {\bf R}$ gives the degree 
of nonextensivity, $k$ is a positive constant.  
Without loss of  generality, we employ $k=1$ in the 
following analysis.
By using the above definitions with $Tr \rho = 1$ 
one obtains the 
canonical distribution\cite{CT}, 
\begin{equation}
p_{n}=p(E_{n})=\frac{\left[ 1-\left( 1-q\right)
 \beta E_{n}\right]^{1/\left( 1-q\right) }}{Z_{q}},  \label{F3}
\end{equation}
where $\{ p_n \}$ are the probabilities, 
$\beta $ is the inverse of the temperature, 
$\left\{ E_{n}\right\} $ is the set of eigenvalues of the
 Hamiltonian, and \begin{equation} 
Z_q=\sum_n\left[ 1-(1-q)\beta 
E_n\right]^{1/\left( 1-q\right) }   
\label{F4}
\end{equation}
is the generalized partition function. 
In the expressions 
(\ref{F3}) and (\ref{F4}) we assumed that $1-\left( 1-q\right)
 \beta E_n\geq 0$. When this condition is not satisfied 
we have a cut-off. 
For instance, when a classical partition function is 
calculated, the integration 
limits in phase space are given by the condition 
$1-\left( 1-q\right) \beta H\geq 0$.
>From Eqs. (\ref{F3}) and (\ref{F4}) several relations can
 be obtained, for instance, 
the generalized free energy becomes 
\begin{equation}
F_q=U_q-TS_q=-\frac 1\beta \frac{Z_q^{1-q}-1}{1-q},  \label{F5}
\end{equation}
where $U_q=\sum_np_n^qE_n$ is the generalized internal energy, 
and $T$ is the temperature which satisfies the relation 
$1/T=\partial S_q/\partial U_q$. 
Note that the previous expressions are reduced to the usual 
ones in the limit case $q\rightarrow 1$.


To develop the perturbation method in this 
generalized statistical mechanics we assume that the Hamiltonian 
of the system is 
\begin{equation}
H=H_0+\lambda H_I. 
\label{F6}\end{equation}
In this expression, $H_{0}$ is the 
Hamiltonian of a soluble model, $\lambda H_I$ is small 
enough so that it can be 
considered as a perturbation on $H_0$ 
($H_0$ and $H_I$ need not necessarily commute), and $\lambda $ 
is the perturbation parameter. 
Thus, the perturbative expansion 
of the free energy can be written as 
$F_q(\lambda )=F_q^{(0)}+\lambda F_q^{(1)}+
\frac{\lambda ^2}2F_q^{(2)}+ \ldots \; .$

To understand how  the 
corrections $F_q^{(n)}$ are calculated it  suffices to 
evaluate the first three terms. 
The first 
term is the free energy for the case without perturbation, 
$F_q^{(0)}=F_q(0)$. 
The second term is obtained from the first derivative
 of $F_q(\lambda )$ 
at $\lambda =0$, 
\begin{equation}
F_q^{(1)}=\frac{\partial F_q\left( 0\right) }
{\partial \lambda }=\left. -
\frac 1{\beta Z_q^q}\frac \partial {\partial \lambda }
\sum_n\left[ 1-\left(1-q\right) 
\beta E_n\right] ^{1/\left( 1-q\right) }
\right| _{\lambda=0}=\langle H_I\rangle _q^{\left( 0\right) }, 
 \label{F8}
\end{equation}
where the superscript $\left( 0\right) $ 
indicates that the $q$-expected value is calculated
 for $\lambda =0$. 
To obtain the last equality it is necessary 
to exchange the order of derivative with 
respect to $\lambda$ and the sum over $n$. 
Furthermore,   the Hellmann-Feymann theorem was used ,
 $\partial E_{n}/ \partial \lambda = \langle n| \partial H/
 \partial \lambda |n \rangle$, or 
equivalently the relation  
$\partial \mid n\rangle /
\partial \lambda=\sum_{m\neq n}\langle m\left|
 H_I\right| n\rangle /
\left( E_n-E_m\right) $, which was  
obtained from the first order perturbation theory. 
The calculation of $F_q^{(2)}$ is similar.
Thus, by using the above considerations we obtain  
\begin{eqnarray}
F_q^{(2)} &=&\frac{\partial ^2F_q\left( 0\right) }
{\partial \lambda ^2} \nonumber \\
&=&
-\beta q \left( Z_q^{\left( 0\right)} 
\right)^{q-1}
 \sum _{n} p(E_n^{(0)}) \left[ \left( p(E_n^{(0)})^{q-1}
\langle n|H_I|n\rangle^{(0)} -
 \langle H_I\rangle_{q}^{(0)} \right) ^2
\right]\nonumber \\
&-&\sum_n\sum_{m\neq n}\left| \langle n
\left| H_I\right| m\rangle^{(0)} \right| ^2 
\frac{p\left( E_m^{(0)}\right) ^q-p
\left( E_n^{(0)}\right) ^q}{E_n^{\left(0\right) }-E_m^{(0)}} \; .
\label{F9}
\end{eqnarray}
When $H_0$ and $H_I$ commute (in the classical case, for instance) 
the expression (\ref{F9}) is more simple, 
i. e. the second term in the right side is zero. 
Notice  that the Eqs. (\ref{F8}) and (\ref{F9}) are correct 
for any $\lambda $, 
but in this case it is necessary to consider the dependence of 
$\mid n\rangle $ with $\lambda $ and to 
substitute $E_n^{(0)}$ by $E_n$. 
This property will be used later on.

Finally, the free energy up to the corrections 
calculated above is 
\begin{eqnarray}
F_q\left( \lambda \right) &=&F_q\left( 0\right) 
+\lambda \langle H_I\rangle_q^{\left( 0\right) }
-\frac{\lambda ^2}2\beta q 
\left(Z_q^{\left( 0\right)}\right)^{q-1}
\sum _{n} p(E_n^{(0)}) \left[ \left( p(E_n^{(0)})^{q-1}
\langle n|H_I|n\rangle^{(0)} -
 \langle H_I\rangle_{q}^{(0)} \right) ^2
\right]\nonumber \\
&-&\sum_n\sum_{m\neq n}\left| \langle n
\left| H_I\right| m\rangle^{(0)} \right| ^2
 \frac{p\left( E_m^{(0)}\right) ^q-p
\left( E_n^{(0)}\right) ^q}{E_n^{\left(0\right) }-E_m^{(0)}}
+{\cal O}\left( \lambda ^3\right) ,  
\label{F11}
\end{eqnarray}
where $\mid n\rangle $ is evaluated with $\lambda =0$. 
It is important to emphasize that the previous calculation was 
performed supposing that the derivative with 
respect to $\lambda $ can commute  
with the sum over the states, but generally it is not true 
 because 
the sum, through its upper limit, can depend on $\lambda $. 
Indeed, when the factor 
$1-\left( 1-q\right) \beta E_n$ is negative the 
sum in(\ref{F4}) 
must be truncated, and as 
$E_n$ depends on $\lambda $ it follows that the upper limit
 of the sum over 
the states also depends  on $\lambda $. 
Without losing   generality one can 
consider $\beta >0$ in the following analysis 
(for $\beta <0$ the analysis 
is essentially the same). 
Therefore, we  can suppose that $E_n\geq 0$ for all $n$. 
In this case, $1-\left( 1-q\right)\beta E_n$ is not 
negative for $q>1$, 
then the order of the derivative in $\lambda $ with 
the sum in $n$ can be interchanged. 
For the case $q<1$ the analysis is more complicated,
 and we will return 
to this question later in the classical context.


Consider now the variational method. 
In this case, the Hamiltonian of the system is $H=H_0+H_{I}$, 
but in the following analysis it is convenient
 to employ a Hamiltonian 
that interpolates continuously $H_0$ and $H $. 
Thus, we use the Hamiltonian 
(\ref{F6}) with $\lambda \in \left[0,1\right] $. 
Now, we will employ the general identity for  
free energy, 
\begin{equation}
F_q\left( \lambda \right) =F_q\left( 0\right) 
+\lambda F_q^{\prime }\left(0\right) 
+\frac{\lambda ^2}2F_q^{^{\prime \prime }}
\left( \lambda _0\right) ,
\label{F12}
\end{equation}
where each prime indicates a derivative 
with respect to $\lambda $, and
 $\lambda _0\in \left[ 0,1\right] $. 
On the other hand, the expression (\ref{F9})
 is not positive for $q>0$, 
and it is not negative for $q<0$ for $\lambda = \lambda_0$. 
In fact, $\sum _{n} p(E_n) \left[ \left( p(E_n)^{q-1}
\langle n|H_I|n\rangle - \langle H_I\rangle_{q} \right) ^2
\right] \geq 0 $, and $p\left( E_n\right) 
\leq p\left( E_m\right) $ for $E_n \geq E_m$ .
Therefore, by considering (\ref{F8}) and (\ref{F12})  
we conclude that $F_q\left(\lambda \right) 
\geq F_q\left( 0\right) 
+\lambda \langle H_I\rangle _q^{(0)}$ for $q<0$,
 and $F_q\left( \lambda \right) 
\leq F_q\left( 0\right) +\lambda\langle H_I\rangle_q^{(0)}$ 
for $q>0$. 
But these conclusions, as we discussed at the end of 
the last paragraph, 
cannot always be true for all $q$. 
Thus, when we consider $\beta >0$, $E_n\geq 0$ 
and the fact that the last inequality is correct for all 
possible values of $\lambda $, 
we can choose $\lambda =1$ and conclude that 
\begin{equation}
F_q\leq F_q^{\left( 0\right)} 
+\langle H_I\rangle _q^{\left( 0\right) }
\label{F13}
\end{equation}
for $q\geq 1$ . 
As we can see immediately,  
the inequality (\ref{F13}) is a natural generalization of
 the 
usual Bogoliubov inequality, $F {\leq} F^{\left(0\right)} 
+\langle H_I\rangle ^{\left( 0\right) }$.


To analyse  the contribution 
of the cut-off in the perturbation and variational methods 
it is convenient 
to consider the generalized classical statistics 
for $\beta >0$ and 
$E_n\geq 0$ (the analysis for $\beta<0$ and $E_n<0$ 
can be extended analogously). 
In this case, there is  cut-off only for
the case $q<1$, so we are going to restrict the following 
analysis to the case $q<1$.
Now, we change $\sum_n$ by $\int d\Gamma $, where $d\Gamma =
\prod_s\left( dx_sdp_s/h\right) $ ($h$ is the Planck constant), 
and the integration is over the phase space region 
defined by the 
inequality 
$1-\left( 1-q\right) \beta H\geq 0$. 
Thus, to analyse $F_q^{(n)}$ we make use of   
the following identity 
\begin{equation}
\frac d{d\lambda }\int d\Gamma f=
\int d\Gamma \frac{\partial f}{\partial\lambda }
+\int_{\partial V}\sum_u dS_u
\left( f\frac{dy_u}{d\lambda }\right) .
\label{F14}
\end{equation} 
In the present application, $f$ is a 
function of phase space variables, $y_u$, 
and $\partial V$ is the hypersurface 
defined by the equation $1-\left(1-q\right) \beta H=0$.

When the last term in the right side 
of (\ref{F14}) is zero, the previous analysis is recovered. 
In general, 
this occurs when $f=0$ and $dy_n/d\lambda $ is 
finite on $\partial V$. 
By using these conditions and 
$f=\left[ 1-\left( 1-q\right)
 \beta H\right]^{1/\left( 1-q\right) }$ 
we conclude immediately that (\ref{F8})
 remains intact for $q<1$. 
For $F_q^{(2)}$ the 
function $f$ contains a term  proportional
 to $\left[ 1-\left( 1-q\right) 
\beta H\right]^{q/\left( 1-q\right) }$. 
Thus, the condition $f=0$ on $\partial V$ 
is satisfied only for $q/ \left(1-q\right) >0$, i. e. $q>0$. 
For $F_q^{(n)}$ the function $f$ contains terms proportional 
to $\left[ 1-\left( 1-q\right) 
\beta H\right]^{\left[nq-\left( n-1\right) \right] /
\left( 1-q\right) }$. 
Consequently, it is necessary $q>1-1/n$ in order that $F_q^{(n)}$ 
does not  contribute to the 
last term of (\ref{F14}). 
These observations indicate that  arbitrarily high orders in the 
perturbation expansion, developed in preceding discussions,  
can be used only for $q\geq 1$. 
On the other hand, the variational approach can be used, 
in general ,  for all $q>0$. 
When the last term of (\ref{F14}) is not zero, there 
is no guarantee that 
$F_q^{(2)}$ remains negative, because, in general, 
the sign of $dy_u/d\lambda $  is not known. 
In this case, the inequality (\ref{F13}) 
cannot be generally employed.


To illustrate the  methods developed above, we consider a 
one-dimensional harmonic oscillator, 
$H=p^{2}/2m+mw^{2}x^{2}/2$, which we will approximate 
by a particle in a square well potential, $H_{0}=p^{2}/2m+V_{0}$ 
with $V_{0}=0$ for 
$ \left | x \right| \leq L/2$ and 
$V_{0}=\infty $ for $\left|x\right| > L/2$. 
In this example, we are 
assuming $\beta>0$.  
The partition 
function of the unperturbed system  and the $q$-expectation 
value for 
$H_{1}=H-H_{0}$ 
can be directly calculated, i. e.
\begin{eqnarray}
Z_{q}^{(0)}= \left \{ \matrix{
\frac{L}{h}\left[ \frac{2m \pi}{(1-q)\beta }\right] ^{\frac{1}{2}}
\frac{\Gamma 
\left( \frac{2-q}{1-q}\right) }{\Gamma 
\left( \frac{2-q}{1-q}+\frac{1}{2}\right) } \; ,
&  q<1 \cr 
\frac{L}{h}\left[ \frac{2m \pi }{(q-1)\beta }\right] ^{\frac{1}{2}}
\frac{\Gamma 
\left( \frac{1}{q-1} - {1\over 2} \right)}{\Gamma 
\left( \frac{1}{q-1}\right) } \; ,
&   q>1 \cr} \right.  
\label{X1}
\end{eqnarray}
and
\begin{eqnarray}
\langle H-H_{0} \rangle^{(0)}_{q}= \left \{ \matrix{
\frac{m \omega^{2}L^{3}}{24hZ_{q}^{(0) ^q}}
\left[ \frac{2m \pi}{(1-q)
\beta }\right]^{\frac{1}{2}}\frac{\Gamma 
\left( \frac{1}{1-q}\right)}
{\Gamma\left( \frac{1}{1-q}+\frac{1}{2}\right) } \; ,
&  q<1 \cr 
\frac{m \omega^{2}L^{3}}{24hZ_{q}^{(0)^q}}
\left[ \frac{2m \pi}{(q-1)
\beta }\right]^{\frac{1}{2}}
\frac{\Gamma \left( \frac{q}{q-1} -{1\over 2} \right)}
{\Gamma\left( \frac{q}{q-1}\right) } \; ,
&  q>1 \cr} \right. .  \label{X2}
\end{eqnarray}
Thus, the free energy, $F_{q}(0)$, can be obtained directly 
from Eqs. (\ref{X1}) and (\ref{F5}).
Note that, $Z_{q}^{(0)}$ is  convergent only for $q < 3$.

Now, we consider the variational method. 
In this case, the minimization of 
$ F_{q}^{(0)}+\langle H-H_{0} \rangle_{q}^{(0)}$ leads to 
\begin{eqnarray}  
\label{A13}
L=\frac{4}{3-q} \left(\frac{3}{m \omega^2 \beta}\right)^{1/2} \; 
\end{eqnarray}
for both $q<1$ and $q>1$. 
Substitution of this result in 
$ F_{q}^{(0)}+\langle H-H_{0} \rangle_{q}^{(0)}$ leads to the
optimum approximation for the free energy. 
Finally, the comparison of this approximation for the free energy 
with the exact one, (valid only for $q<2$) 
\begin{equation}
F=-\frac{1}{(1-q)\beta }
\left[ \left( \frac{2\pi}{h \omega \beta} 
\frac{1}{2-q} \right )^{1-q}-1\right] \;, 
\end{equation}
gives a good approximation, as in the case 
$q \rightarrow 1$(see fig.
(\ref{fig1}) ). Furthermore,  fig. (\ref{fig1}) shows that the
approximation is improved for larger $q$.
Notice that all previous expressions reduce to the usual one
in the limit $q \rightarrow 1$. 
The perturbative contributions can be obtained in a similar way.


Summing up, we have developed here generalized 
perturbation and variational methods for the nonextensive 
context in a unified way. 
In this approach, a generalization of the Bogoliubov 
inequality which is form-invariant for all $q$ is obtained.  
This property is in variance with the generalization
presented in ref.\cite{PT}. 
When we consider arbitrary high orders in 
the perturbation expansion, we must have  $q\geq 1$. 
On the other hand, the Bogoliubov inequality can be generally used 
for $q>0$.
We believe that  approaches presented here are useful 
in the discussion
of the anomalies currently associated with nonextensive systems.

\acknowledgements
We  gratefully acknowledge  useful remarks by A. K. Rajagopal.
We also thank partial finantial support by CNPq (Brazilian Agency).


\begin{figure}
\caption{Free energy approximated and exact vs.
 temperature for three typical values of $q$ (with $\omega=m=h=1$).}
\label{fig1}
\end{figure}


\references
\bibitem {1a} 
M. F. Shlesinger, B. J. West and  J. Klafter, 
{ Phys. Rev. Lett.}  {\bf 58}, 1100 (1987).
\bibitem {1b}
J. P. Bouchaud  and A. Georges, 
{ Phys. Rep.} {\bf 195}, 127 (1991).
\bibitem {1c}
M. F. Shlesinger, G. M. Zaslavsky and J. Klafter, 
{ Nature}  {\bf 363}, 31 (1993).
\bibitem {1d}
J. Klafter, G. Zumofen and A. Blumen,   
{ Chem. Phys.}  {\bf 177}, 821 (1993).
\bibitem {2}
  A. M. Salzberg  { J. Math. Phys.} {\bf 6}, 158 (1965).
\bibitem {2a}L. Tisza, {\it Generalized Thermodynamics} 
(MIT Press, Cambridge,1966) p. 123.
\bibitem {2b}
P. T. Landsberg,  
{ J. Stat. Phys.} {\bf 35}, 159 (1984).  
\bibitem {2c}
J. Binney  and  S. Tremaine,  
{\it Galactic Dynamics} (Princeton University Press, 
Princeton, 1987) p. 267. \bibitem {2d}
H. S. Robertson, {\it Statistical Thermophysics}  
(P. T. R. Prentice-Hall, 
Englewood Cliffs, New Jersey 1993) p. 96.
\bibitem {3}
 B. J. Hiley  and G. S. Joyce,  
{ Proc. Phys. Soc.} {\bf 85}, 493 (1965). 
\bibitem {3a}
S. K. Ma,   {\it Statistical Mechanics} 
(World Scientific, New York, 1993) p. 116. 
\bibitem {3b}
S. A. Cannas,  
{ Phys. Rev.} { B}   {\bf 52}, 3034 (1995). 
\bibitem {4}
 J. O. Indekeu, {Physica} A {\bf  183}, 439  (1992). 
\bibitem {4a}
J. O. Indekeu and A. Robledo, 
{ Phys. Rev.}  E {\bf 47}, 4607 (1993).
\bibitem {T1}
C. Tsallis, {J. Stat. Phys.} {\bf 52}, 479 (1988).
\bibitem {6}
C. Tsallis, S. V. F. Levy, A. M. C. Souza  and R. Maynard,  
{ Phys. Rev. Lett. } {\bf 75}, 3589 (1995); 
Erratum: {\bf 77}, 5442 (1996);
D. H. Zanette andd P. A. Alemany, {Phys. Rev. Lett.}
 {\bf 75}, 366 (1995);
M. O. Caceres and C. E. Budde, {Phys. Rev. Lett.}
 {\bf 77}, 2589 (1996).  
\bibitem {7}
 B. M. Boghosian,  { Phys. Rev.} E {\bf 53}, 4754 (1996). 
\bibitem {8}
 A. R. Plastino  and  A. Plastino,    
{ Phys. Lett.} A {\bf 174}, 384 (1993);
J. J. Aly,  {\it Proceedings of N-Body Problems 
and Gravitational Dynamics, Aussois, France} ed
 F. Combes and E, Athanassoula 
(Publications de l'Observatoire de Paris, Paris, 1993) p. 19.
\bibitem {8a}
V. H. Hamity and D. E. Barraco, {Phys. Rev. Lett.}
 {\bf 76}, 4664 (1996).
\bibitem {8b} 
L. P. Chimento, J. Math. Phys. {\bf 38}, 2565 (1997).
\bibitem {8c}
D. F. Torres, H. Vucetich and A. Plastino, {Phys. Rev. Lett.} 
(in press).
\bibitem {9}
 C. Tsallis, F. C. S\'a Barreto  and E. D. Loh, 
 {Phys. Rev.} E {\bf 52}, 1447 (1995). 
\bibitem {10a}
 A. Lavagno, G. Kaniadakis, M. Rego-Monteiro, P. Quarati and
C. Tsallis, {Astro. Lett. and Comm.} (1997), in press.
\bibitem {10b}
A. K. Rajagopal, {Phys. Rev. Lett.} {\bf 76}, 3469 (1996).
\bibitem {10c}
I. Koponen, { Phys. Rev.} {E} {\bf 6}, 7759 (1997).
\bibitem {11}
 P. Jund, S. G. Kim  and C. Tsallis,  {Phys. Rev.} B {\bf 52}, 50 (1995).
\bibitem {L}
B. H. Lavenda and J. Dunning-Davies,
 {Found. Phys. Lett.} {\bf 3}, 435 (1990);
B. H. Lavenda and J. Dunning-Davies, {Nature} 
{\bf 368}, 284 (1994); 
B. H. Lavenda, J. Dunning-Davies and M. Compiani, 
{Nuovo Cimento B} {\bf 110}, 
433 (1995); 
see also  B. H. Lavenda, {\it Statistical Physics:
 A Probabilistic 
Approach}, (Wiley-Interscience, New York, 1991) 
and B. H. Lavenda, {\it Thermodynamics of Extremes}, 
(Albion, Chichester, England, 1995).
\bibitem {L2}
One of the nonextensive systems recently studied
 in the literature is
the blackhole, whose entropy is a controversial subject 
(see ref. \cite{L} and references therein).
This problem needs a separate investigation when 
considered in terms of Tsallis
approach and  lies outside the scope of the present letter.
\bibitem {M}
R. S. Mendes, {Physica A } {\bf 242}, 299 (1997).
\bibitem {EMM}
L. R. Evangelista, L. C. Malacarne and R. S. Mendes, 
{\it Quantum Corrections for General Partition Functions},
 preprint (1997). 
\bibitem {F1}
R. P. Feynman, {\it Statistical Mechanics: 
A Set of Lectures}, 
(W. A. Benjaminm Inc, Massachusetts, 1972) p. 67-71.
\bibitem {PT} 
A. Plastino and C. Tsallis, {J. Phys. A} {\bf 26}, L893 (1993).
\bibitem {CT}  
E. M. F. Curado  and  C. Tsallis,  {J. Phys. A} {\bf 24}, L69 (1991);
Errata: {\bf 24}, 3187 (1991); {\bf 25}, 1019 (1992).
\end{document}